\documentclass{JINST}
\usepackage{epsfig}
\usepackage{subfigure}
\usepackage{here}
\usepackage{graphicx}

\def\figurename{\bf{Figure\space}}

\title{Further progress in ion back-flow reduction with patterned gaseous hole - multipliers.}

\author{A. V. Lyashenko$^a$\thanks{Corresponding author.}, A. Breskin$^a$, R. Chechik$^a$, J. F. C. A. Veloso$^b$, J. M. F. Dos Santos$^c$, and F. D. Amaro$^c$\\
\llap{$^a$} Department of Particle Physics, Weizmann Institute of Science,\\
  76100, Rehovot, Israel\\
\llap{$^b$}University of Aveiro,\\
  3810-193 Aveiro, Portugal\\
\llap{$^c$}University of Coimbra,\\
  3004-516 Coimbra,  Portugal\\
E-mail: \email{alexey.lyashenko@weizmann.ac.il}}

\abstract{A new idea on electrostatic deviation and capture of
back-drifting avalanche-ions in cascaded gaseous hole-multipliers is
presented. It involves a flipped reversed-bias Micro-Hole \& Strip
Plate (F-R-MHSP) element, the strips of which are facing the drift
region of the multiplier. The ions, originating from successive
multiplication stages, are efficiently deviated and captured by such
electrode. Experimental results are provided comparing the
ion-blocking capability of the F-R-MHSP to that of the reversed-bias
Micro-Hole \& Strip Plate (R-MHSP) and the Gas Electron Multiplier
(GEM). Best ion-blocking results in cascaded hole-multipliers were
reached with a detector having the F-R-MHSP as the first
multiplication element. A three-element F-R-MHSP/GEM/MHSP cascaded
multiplier operated in atmospheric-pressure Ar/CH$_{4}$ (95/5), at
total gain of $\sim$10$^{5}$, yielded ion back-flow fractions  of
3$\cdot$10$^{-4}$ and 1.5$\cdot$10$^{-4}$, at drift fields of 0.5
and 0.2 kV/cm, respectively. We describe the F-R-MHSP concept and
the relevance of the obtained ion back-flow fractions to various
applications; further ideas are also discussed.}

\keywords{ electron multipliers (gas); avalanche induced secondary
effects; charge transport and multiplication in gas; detector
modeling and simulations II}

\begin{document}

\section{Introduction.}

In a recent article \cite{lyashenko:06} we summarized the state of
the art in avalanche-ion blocking in cascaded gas-avalanche electron
multipliers combining GEM and Microhole Hole \& Strip (MHSP)
\cite{veloso:00} elements. The particular cases of Gaseous
Photomultipliers (GPM) and Time Projection Chambers (TPC) were
discussed in some detail. With some of the techniques discussed in
\cite{lyashenko:06}, the ion backflow fraction (IBF), namely the
fraction of the final-avalanche ions flowing back to the drift
volume of a tracking detector, or that impinging on the photocathode
(PC) surface of a GPM, could be reduced to values below 10$^{-3}$ in
a direct current (DC) operation mode. We discussed the fact that
this number is still too large in some applications and that a
gated-mode operation, though reducing the IBF down to values below
10$^{-4}$ \cite{mormann:04, breskin:05}, cannot always be a solution
(lack of trigger, dead-time, pick-up noise etc.). The recently
proposed and currently investigated ion blocking concept by
photon-assisted avalanche propagation in cascaded multipliers
\cite{veloso:06, buzulutskov:06} has some interesting aspects.
However, it is limited to an operation with highly scintillating gas
mixtures, in the far UV; furthermore, methods have to be conceived
for blocking the ions originating from the avalanche developing in
the first multiplication/scintillation element of the cascaded
detector.

\begin{figure} [h]
  \begin{center}
  \makeatletter
    \renewcommand{\p@figure}{figure\space}
  \makeatother
    \epsfig{file=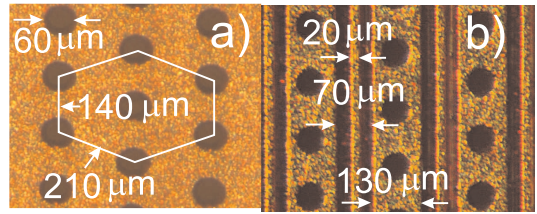, width=7cm}
    \caption{A two-sided microscope photograph of a MHSP electrode with 20$\mu$m anode strips and 130$\mu$m cathode strips.}
    \label{figure:1}
  \end{center}
\end{figure}

Some of the ideas and results published in \cite{lyashenko:06,
veloso:05, roth:04} suggested ion blocking with a reversed-bias
Micro-Hole \& Strip Plate (R-MHSP) used as a first element in a
multipliers' cascade. This method, in which ions are deviated and
neutralized on strip-electrodes patterned on the surface of a
hole-multiplier \ref{figure:1}, provided a good suppression of ions
flowing back from successive elements into the R-MHSP. However, a
large fraction of ions originated from avalanches occurring within
the R-MHSP holes remained unblocked. Therefore, another DC
ion-blocking method have been recently conceived and evaluated,
aiming at more efficient electrostatic deviation of the drifting
ions towards collecting strip-electrodes patterned at the holes
vicinity.

The present work describes the new concept of "Flipped"
Reversed-bias Micro-Hole \& Strip Plate (F-R-MHSP) electrodes in
which strips are pointing up towards the drift region of the
multiplier. Simulation and experimental results of electron
collection and ion blocking of the F-R-MHSP are compared with that
of R-MHSP and GEM. The F-R-MHSP permits good electron collection and
efficient blocking of ions, both originated form the first element
itself and that from avalanches in successive cascade elements. The
IBF reduction in cascaded gaseous micro-hole multipliers
incorporating first-element F-R-MHSPs is demonstrated and discussed
in view of their potential applications in tracking devices and
GPMs.

\section{The F-R-MHSP concept}

\begin{figure}[h]%
\begin{center}%
\renewcommand{\thesubfigure}{\thefigure\alph{subfigure}}
\makeatletter
\renewcommand{\p@subfigure}{figure\space}
\makeatother %
\subfigure {
    \label{figure:2:a}
    \includegraphics[width=7.4cm]{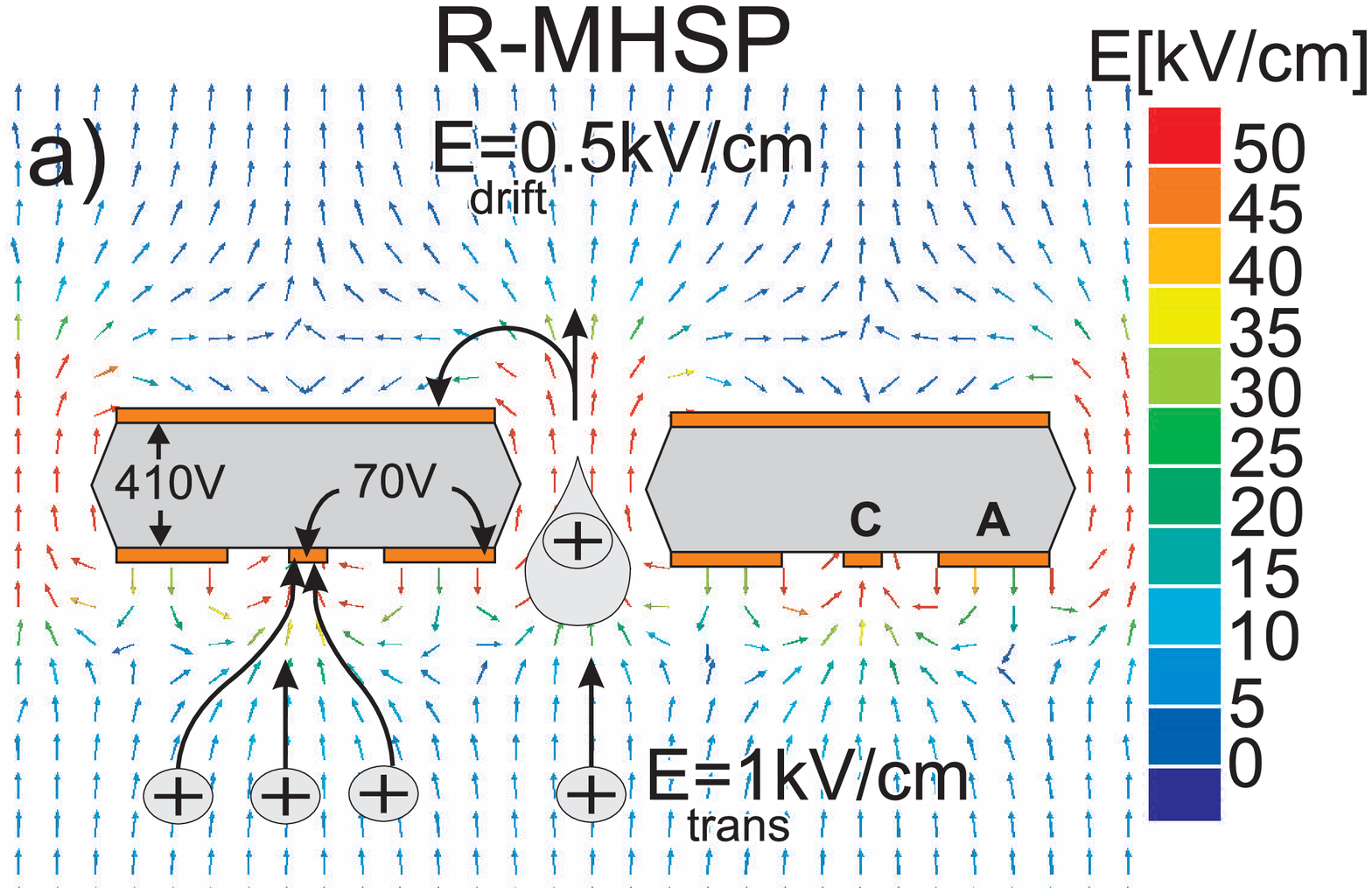}
} \hspace{0cm} \subfigure {
    \label{figure:2:b}
    \includegraphics[width=6.8cm]{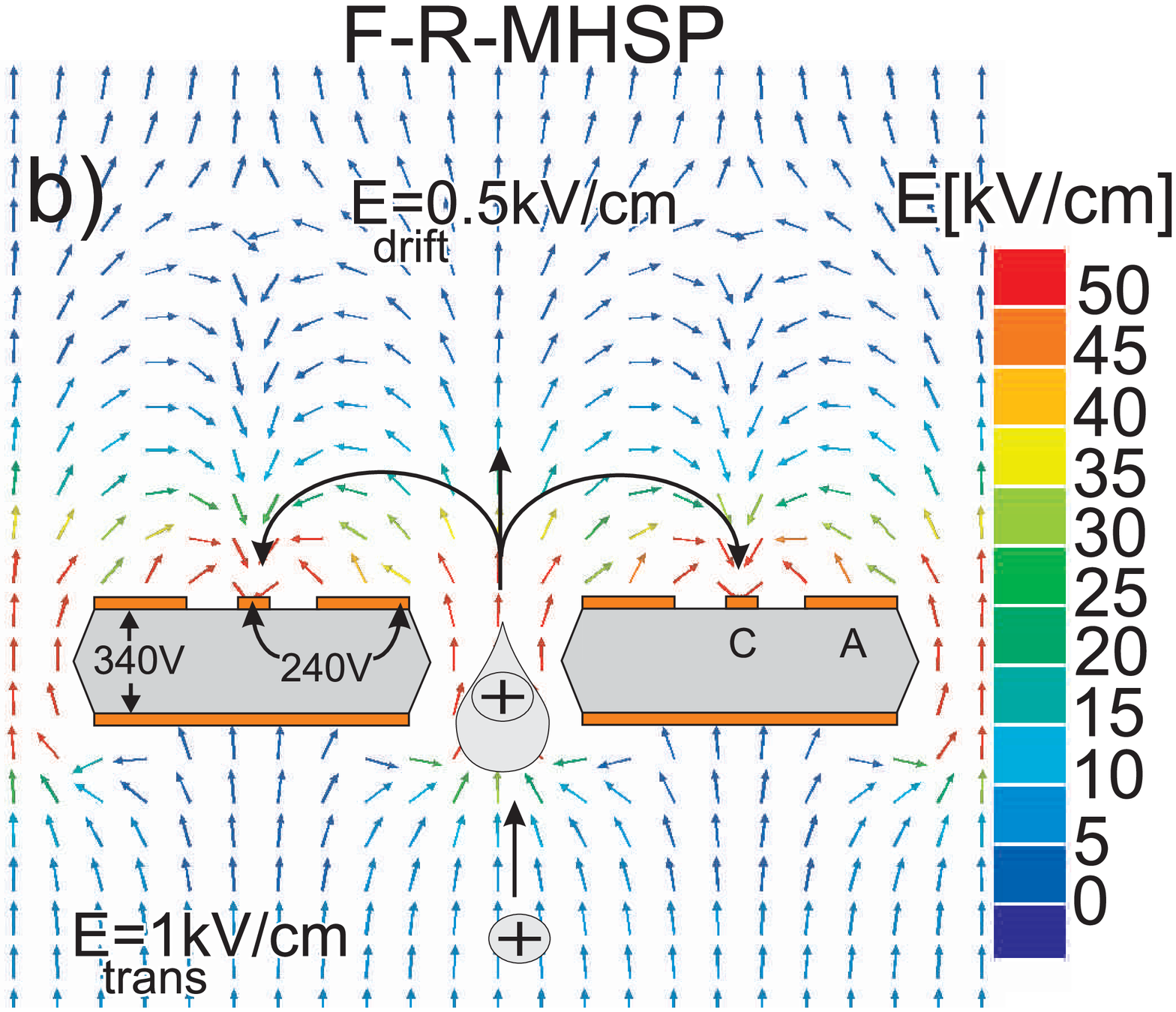}
} \caption{The electric-field vectorial maps calculated by MAXWELL
software package \protect \cite{maxwell} in the vicinity of the
electrodes and schematic views of the operation principles for: a)
reversed-biased R-MHSP and b) flipped reversed biased F-R-MHSP. The
potentials selected for the field-map calculations and the color
code of the fields are shown in the figures.}%caption for the whole figure
\label{figure:2}
\end{center}
\end{figure}

In the R-MHSP (\ref{figure:2:a}), a MHSP electrode is mounted with
its strips facing the next multiplying element; the bias scheme on
the strips is reversed compared to the MHSP, namely the narrow
strips are biased negatively compared to the wide strips
encompassing the holes. Electrons collected into the holes induce
charge multiplication within the holes; unlike the MHSP, there is no
further charge multiplication on the strips.

The negatively charged strips trap ions originated from successive
multiplying elements. However, like in a GEM, avalanche-ions created
within the R-MHSP holes remain uncollected and reach the drift
volume (or the PC). The idea of flipping the R-MHSP with its
patterned strips pointing towards the drift volume (or the PC), aims
at collecting all back-drifting ions, including those originating
from its own hole-avalanches. The flipped R-MHSP (F-R-MHSP) is shown
schematically in \ref{figure:2:b}.  Like in the R-MHSP mode, the
narrow ion-collection strips (\ref{figure:1}) are biased more
negative than the broader ones; the multiplication occurs only
within the holes. The F-R-MHSP can be inserted anywhere along the
cascaded multiplier, but its best performance is expected when used
as the first multiplying element.

\section{Methodology.}

The MHSP and GEM electrodes employed in this work, of 28x28mm$^{2}$
effective area, were produced at the CERN printed circuit workshop,
from 50$\mu$m thick Kapton foil with 5$\mu$m copper cladding on both
sides. The etched double-conical 70/50$\mu$m (outer/inner) diameter
GEM holes are arranged in hexagonal pattern of pitch 140$\mu$m. The
MHSP pattern and dimensions are shown in \ref{figure:1}. All
electrodes were stretched onto small G-10 frames. The
semitransparent PC was 5mm in diameter; it was made of 300{\AA}
thick layer of CsI, evaporated on a UV transparent window,
pre-coated with a 40{\AA} thick Cr film.

The detector elements were mounted within a stainless-steel vessel
evacuated with a turbo-molecular pump to 10$^{-5}$ Torr prior to gas
filling. The detector was operated with Ar/CH$_{4}$ (95/5) at 760
Torr, under regulated gas flow. It was irradiated with a continuous
Ar(Hg) UV-lamp through the window. Each of its electrodes was biased
independently with a CAEN N471A or CAEN N126 power supply.

In all multiplier-cascade configurations, the currents on biased
electrodes were recorded as a voltage-drop on a 40M$\Omega$ resistor
with a Fluke 175 voltmeter of 10M$\Omega$ internal impedance. Their
combined resistance was 8M$\Omega$ from which the anode current was
calculated. The final avalanche-induced currents following charge
multiplication were always kept well below 100 nA by attenuating the
UV-lamp photon flux with absorbers, if necessary, to avoid
charging-up effects. The currents on grounded electrodes were
recorded with a Keithley 485 picoamperemeter.

\section{Experimental studies and results.}

\subsection{Single-electron detection efficiency of R-MHSP and
F-R-MHSP}

In cascaded gaseous hole-multipliers (e.g a R-MHSP, a F-R-MHSP or
simply a GEM), to reach full detection efficiency of single
photoelectrons emitted from a photocathode, or of ionization
electrons radiation-induced within the drift volume, two conditions
have to be fulfilled:

\begin{itemize}
\item
The electron's collection efficiency into holes, particularly in the
application to single-photon GPMs, has to be close to unity; this
was indeed confirmed for GEMs \cite{bachmann:99, richter:02} and
more recently for MHSPs (see below), which have slightly more
"opaque" hole geometry (\ref{figure:1});

\item
The visible gain (defined as the number of electrons, per single
initial electron, transferred from a given multiplier element into a
consecutive electrode \cite{lyashenko:06}), of the first element in
the cascade should be large enough to ensure full efficiency of the
event's detection by the following elements.
\end{itemize}

These two conditions are of prime importance, because an electron
lost at the first multiplication element due to inefficient
focusing, insufficient multiplication or inefficient extraction and
transfer into the next multiplication stage, cannot be recovered.
Indeed, it was found that the R-MHSP biasing scheme reduces the
extraction efficiency of the avalanche electrons from the holes
towards the next element in the cascade, thus reducing the visible
gain of this multiplier \cite{lyashenko:06}. In the F-R-MHSP the
strips are facing the PC and could affect photoelectron focusing
into the holes; therefore, the focusing efficiency has to be
carefully determined.

The photon detection efficiency in GPMs, $\varepsilon_{photon}$,
depends on both: the PC's quantum efficiency (QE) and on the
single-photoelectron detection efficiency $\varepsilon_{det}$; it is
defined as:

\begin{equation}
\label{eq:photoeff} \varepsilon_{photon} = \textrm{QE} \cdot
\varepsilon_{det}.
\end{equation}

$\varepsilon_{det}$ depends on many parameters: the detector
geometry, the gas mixture, the electric field conditions, the
multiplier gain, the electronics system etc. Once emitted from the
photocathode surface into the gas, the photoelectron has to be
focused into the first amplifying stage of the detector, namely into
the holes. The mechanism of electron extraction, transfer and
multiplication in cascaded GEMs for hole voltage values exceeding
320V were extensively studied in \cite{richter:02}.

While the operation properties of the MHSP were well established
\cite{maia:04}, those of the R-MHSP, F-R-MHSP and GEM operating at
hole voltages lower than 320V required some more basic study. The
studies of single-electron detection efficiency for the R-MHSP and
the F-R-MHSP were designed to yield better understanding of the role
of the various potentials and of the conditions for reaching minimal
IBF values while keeping minimal electron losses.

The parameters affecting the R-MHSP and F-R-MHSP operation are:

\begin{enumerate}

\item
the hole voltage (V$_{hole}$); it controls the multiplication and
the IBF values of the first multiplying element;

\item
the anode-to-cathode strip voltage ($\Delta$V$_{AC}$); it reduces
the visible gain of a single R-MHSP; it could affect the focusing
properties of the F-R-MHSP and reduce the IBF from successive
elements and from its own avalanches;

\item
the transfer field below the R-MHSP or the F-R-MHSP (E$_{trans}$ in
\ref{figure:2:a} and \ref{figure:2:b}); it could, in principle,
affect both the IBF from successive elements and the visible gain of
the R-MHSP or the F-R-MHSP;

\end{enumerate}

It should be noted that except second condition, similar remarks
also apply to a GEM.

\begin{figure} [t]
  \begin{center}
  \makeatletter
    \renewcommand{\p@figure}{figure\space}
  \makeatother
    \epsfig{file=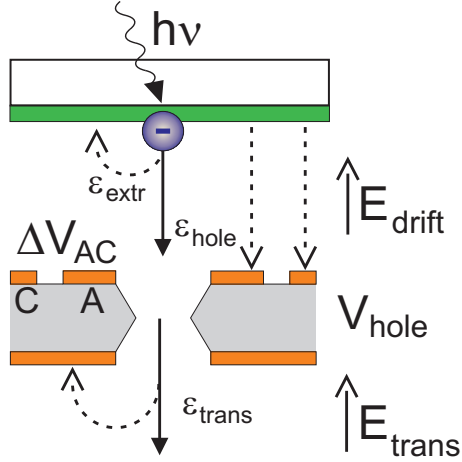, width=6cm}
    \caption{Electron transfer through a F-R-MHSP. Photoelectrons emitted from a
PC are extracted from the photocathode with an efficiency
$\varepsilon_{extr}$, guided into the apertures of the F-R-MHSP with
an efficiency $\varepsilon_{hole}$. A fraction of the avalanche
electrons is extracted and transferred into a following element with
an efficiency $\varepsilon_{trans}$; another fraction is lost to the
bottom electrode. The electron transfer through either a GEM or a
R-MHSP is described the same way, the F-R-MHSP was taken as an
example.}
    \label{figure:3}
  \end{center}
\end{figure}

If we take a GPM as example, the possible fate of the photoelectron
after its emission from the photocathode is schematically shown in
\ref{figure:3}.

The single-photoelectron detection efficiency can thus be described
as:

\begin{equation}
\label{eq:deteff} \varepsilon_{det} = \varepsilon_{extr} \cdot
\varepsilon_{hole} \cdot \varepsilon_{trans}
\end{equation}

Here $\varepsilon_{det}$ is the probability to detect a single
photoelectron, $\varepsilon_{extr}$ is the probability to extract a
photoelectron from the PC, $\varepsilon_{hole}$ is the probability
to get this electron into the hole and $\varepsilon_{trans}$ is the
probability to transfer an avalanche electron to the next
multiplication stage.

As mentioned above, all the measurements were performed in
Ar/CH$_{4}$(95/5) under atmospheric pressure. In all the
measurements presented below we assumed $\varepsilon_{extr}$ = 1 in
equation (\ref{eq:deteff}). Realistic $\varepsilon_{extr}$ values
are well known, and were previously measured as function of the
drift field above a photocathode for a variety of gas mixtures
\cite{buzulutskov:00}. Thus the measured electron detection
efficiency in TPC or GPM conditions (low or high drift fields) can
be straightforwardly corrected, using the known $\varepsilon_{extr}$
at the corresponding drift field and gas mixture. For instance, in
atmospheric Ar/CH$_{4}$ (95/5) used in all our measurements,
$\varepsilon_{extr}$=70\% at drift fields of 0.5kV/cm. In TPC
conditions, there is no PC and the parameter $\varepsilon_{extr}$ is
not relevant; therefore, equation (\ref{eq:deteff}) with
$\varepsilon_{extr}$=1 is exact.

Due to the statistical fluctuations in the amplification process of
single electrons, many events have only a small number of electrons
at the exit of the holes of the first amplifying element. DC
measurements (e.g. ratio of the current after multiplication to that
of primary photoelectrons) are not sensitive to single-photoelectron
losses or to events with small gain; their contribution to the total
current is negligible when the detector is operated in
multiplication mode. Under these conditions, the only way to assess
the single-electron detection efficiency $\varepsilon_{det}$, is by
event pulse-counting. Current-mode measurements provide valid
results for single-photoelectron transport, only if the detector is
operated at unity gain. In these conditions, currents measured on
the detector's electrodes are due to the transfer of the primary
photoelectrons only - which is not a subject of our studies. A more
detailed discussion on this subject can be found in
\cite{mormann_thes}.

\begin{figure}[t]
\begin{center}
\renewcommand{\thesubfigure}{\thefigure\alph{subfigure}}
\makeatletter
\renewcommand{\@thesubfigure}{\Large(\alph{subfigure})}
\renewcommand{\p@subfigure}{figure\space}
\renewcommand{\p@figure}{figure\space}
\makeatother %
\subfigure [] [] % caption for subfigure a
{
    \label{figure:4:a}
    \includegraphics[width=7.3cm]{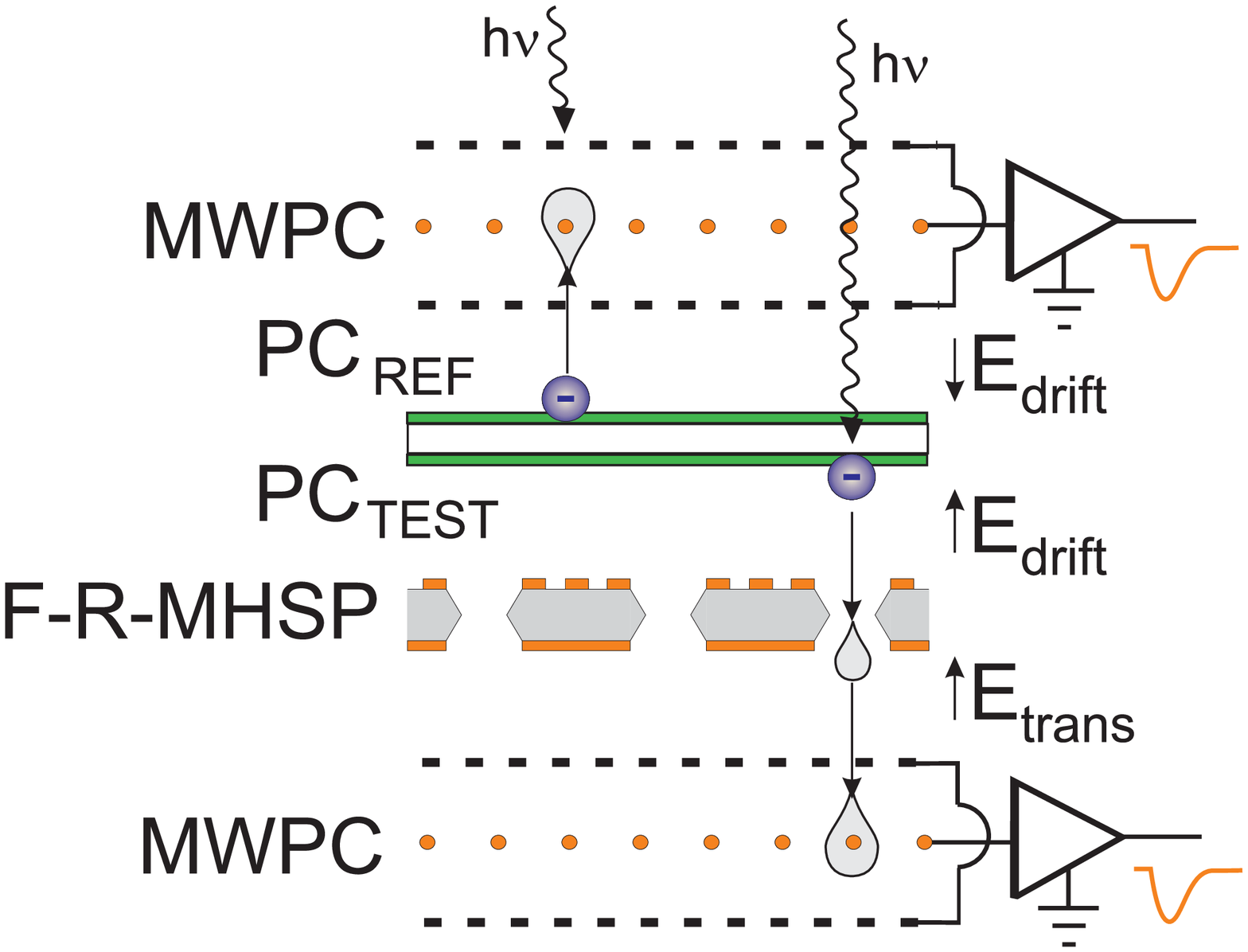}
}\hspace{0.2cm}%
\subfigure [] [] % caption for subfigure b
{
    \label{figure:4:b}
    \includegraphics[width=7cm]{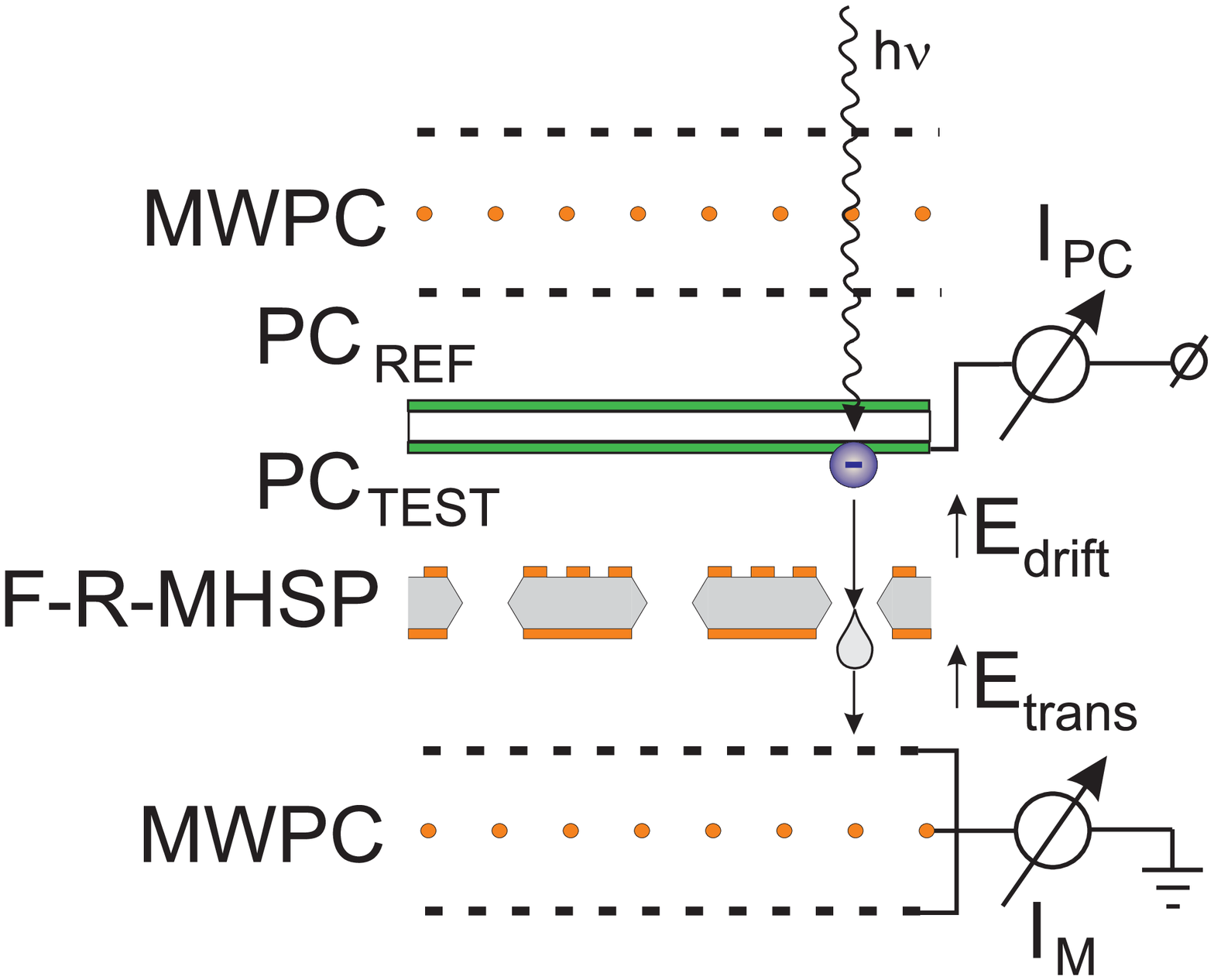}
} \caption {The schematic view of the experimental setup for
measuring the single-electron detection efficiency of a F-R-MHSP (a)
and its visible gain (b). Similar setups were used for measuring the
same properties of  a R-MHSP and a GEM} % caption for the whole figure
\label{figure:4}
\end{center}
\end{figure}

We found it convenient to measure $\varepsilon_{det}$ of the hole
multipliers by comparing their event-rate to the one measured with a
multi-wire proportional counter (MWPC) - known to have
$\varepsilon_{det}$=1 \cite{richter:02, mormann:04a, shalem:06}.
This strategy has been used in many of our previous studies, and was
discussed in detail \cite{richter:02, mormann:04a, shalem:06}.
\ref{figure:4:a} shows the dedicated experimental setup. It consists
of a UV-transparent quartz (Suprasil) window with CsI
semitransparent PCs evaporated on both surfaces, sandwiched between
the reference MWPC detector and the investigated multiplier
(R-MHSP/MWPC, F-R-MHSP/MWPC or GEM/MWPC). Both, reference and
investigated detectors operate at equal total gains; the MWPC
following each investigated hole-multiplier was added to keep the
total gain high enough for pulse counting. The ratio of the number
of detected events, with equal electronic thresholds, multiplied by
the ratio of initial photocurrents of the top-face and bottom-face
PCs, provided the absolute single-electron detection efficiency of
the GEM, the R-MHSP and the F-R-MHSP multipliers. This experimental
technique relies on precise measurements of the exponential
pulse-height distributions of the multiplied single electrons,
needed for adjusting equal conditions in both the reference MWPC and
the investigated hole-multiplier/MWPC elements. This method can be
naturally applied only in proportional-mode operation; it is no
longer applicable in conditions of charge saturation or with
feedback effects - leading to spectra that deviate significantly
from the exponential. A more detailed explanation on this method can
be found elsewhere \cite{mormann:04a,shalem:06}.

\begin{figure}[!h]
\begin{center}
\renewcommand{\thesubfigure}{\thefigure\alph{subfigure}}
\makeatletter
\renewcommand{\@thesubfigure}{\Large(\alph{subfigure})}
\renewcommand{\p@subfigure}{figure\space}
\renewcommand{\p@figure}{figure\space}
\makeatother %
\subfigure [] [] % caption for subfigure a
{
    \label{figure:5:a}
    \includegraphics[width=10cm]{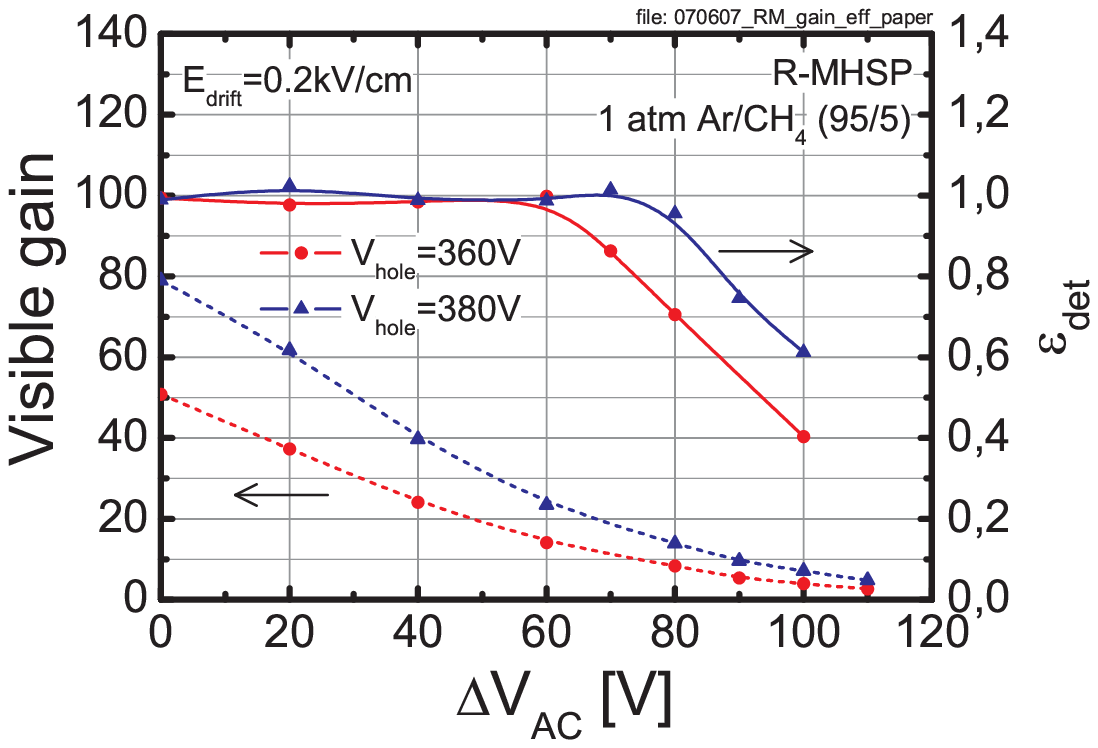}
}\hspace{1cm}%
\subfigure [] [] % caption for subfigure b
{
    \label{figure:5:b}
    \includegraphics[width=10cm]{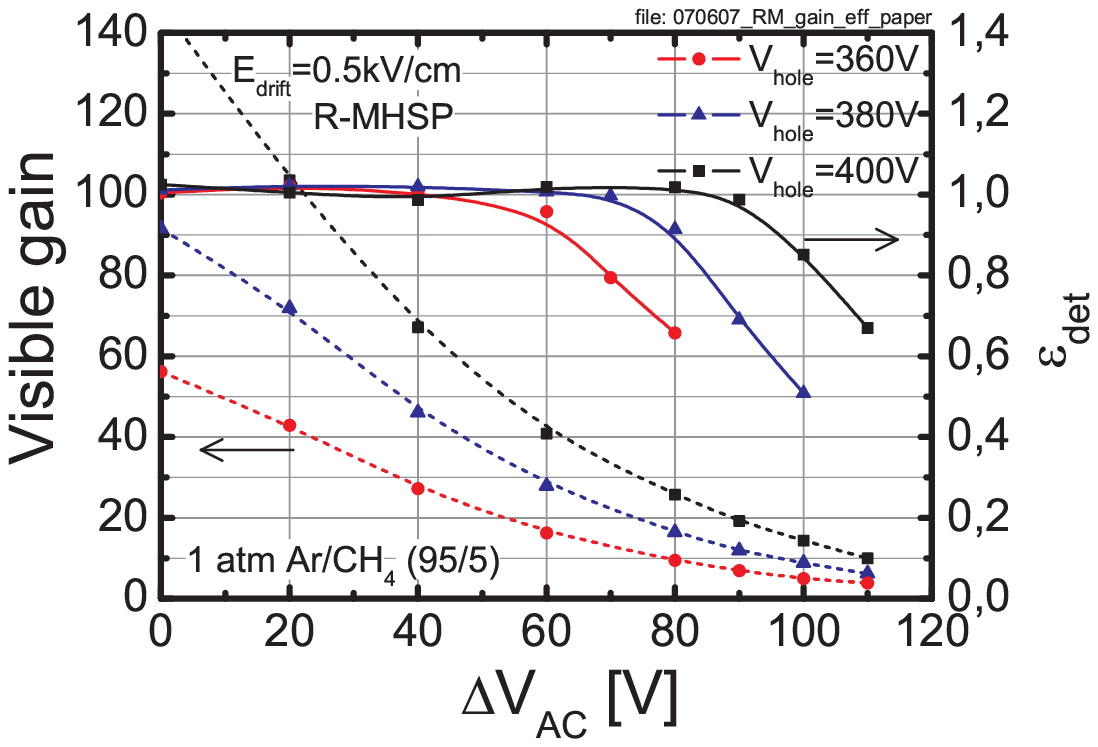}
} \caption {The visible gain (left y-axis, dashed lines) and the
photoelectron detection efficiency (right y-axis, solid lines) as a
function of the inter-strip voltage $\Delta$V$_{AC}$ of a R-MHSP in
TPC conditions (E$_{drift}$=0.2kV/cm) (a) and GPM conditions
(E$_{drift}$=0.5kV/cm) (b). Measurements performed at different values of hole voltage.} % caption for the whole figure
\label{figure:5}
\end{center}
\end{figure}

\textbf{Measurements with R-MHSP.} The photoelectron detection
efficiency of the R-MHSP as function of the inter-strip voltage
(potential difference between the anode and cathode strips)
$\Delta$V$_{AC}$ is presented in \ref{figure:5} for typical "TPC
conditions" (i.e. E$_{drift}$=0.2kV/cm) and for "GPM conditions"
(i.e. E$_{drift}$=0.5kV/cm). In GPM conditions, the measurements
were performed at V$_{hole}$ values of 360V, 380V and 400V; in TPC
conditions, at 360V and 380V. In all cases, the transfer field was
set to 1kV/cm. The visible gains as a function of the inter-strip
voltage are also presented for each set of measurements. The visible
gain G$_{VIS}$ is derived from the ratio of the current I$_{M}$
measured at the interconnected electrodes of bottom MWPC (see
\ref{figure:4:b}), to the PC photocurrent I$_{PC0}$, measured in
photoelectron collection mode (no gain):

\begin{equation}
\label{eq:gvis}
\textrm{G}_{VIS}=\frac{\textrm{I}_{M}}{\textrm{I}_{PC0}}.
\end{equation}

From \ref{figure:5} we learn that:

\begin{enumerate}

\item
Both the visible gain and the photoelectron detection efficiency are
only slightly affected by the drift field. The reason is probably
that the field inside the holes is not affected by the drift field
(within the present conditions).
\item
Although it is desirable to increase $\Delta$V$_{AC}$, as to divert
more ions towards the cathode strips, the drop in the visible gain
(\ref{figure:5}), and consequently in the detection efficiency, sets
a limit to this parameter.$\Delta$V$_{AC}$ can be raised if the loss
of electrons is compensated by a further increase of V$_{hole}$.For
each hole-voltage, the maximal strip voltage could be found at which
the photoelectron detection efficiency is close to unity. The
maximal strip voltages at which the detection efficiency is
$\sim$100\% are 60V, 70V and 90V at corresponding hole voltages of
360V, 380V and 400V. These inter-strip voltages correspond to a
visible gain of about 20 on the R-MHSP as can be seen in
\ref{figure:5}.

\end{enumerate}

\textbf{Measurements with F-R-MHSP} Measurements similar to that of
the R-MHSP were performed for the F-R-MHSP. As mentioned above, in
the F-R-MHSP configuration, the strips are facing towards the drift
region or to the PC. Therefore, the electron transfer to the next
amplification stage is expected to be unaffected by varying the
strip voltage. However, the increase of the inter-strip voltage
difference could, in principle, affect focusing of photo-electrons
into the holes of the F-R-MHSP itself. As in the case of the R-MHSP,
here too we need to optimize the strip- and hole-voltages. The
inter-strip voltage has to be large enough for better ion
collection, while the hole-voltage has to be low enough for reaching
lower IBF values in this element; in addition, the condition of
photoelectron detection efficiency close to unity has to be
fulfilled.

\begin{figure}[!h]
\begin{center}
\renewcommand{\thesubfigure}{\thefigure\alph{subfigure}}
\makeatletter
\renewcommand{\@thesubfigure}{\Large(\alph{subfigure})}
\renewcommand{\p@subfigure}{figure\space}
\renewcommand{\p@figure}{figure\space}
\makeatother %
\subfigure [] [] % caption for subfigure a
{
    \label{figure:6:a}
    \includegraphics[width=10cm]{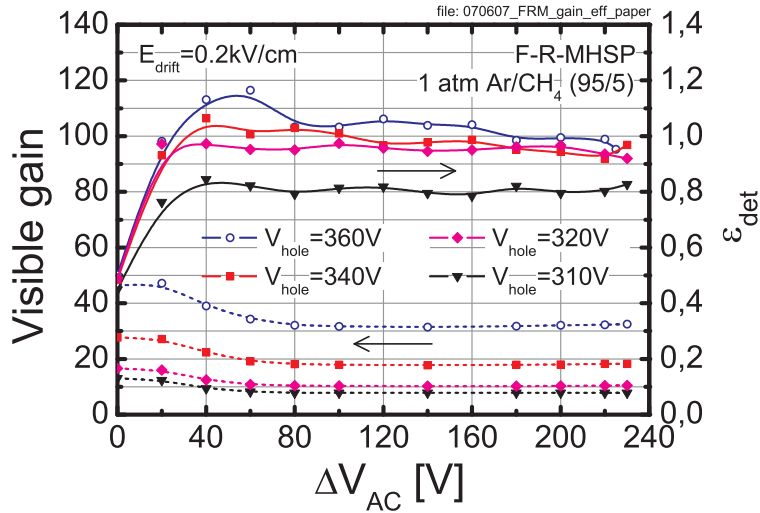}
}\hspace{1cm} %
\subfigure [] [] % caption for subfigure b
{
    \label{figure:6:b}
    \includegraphics[width=10cm]{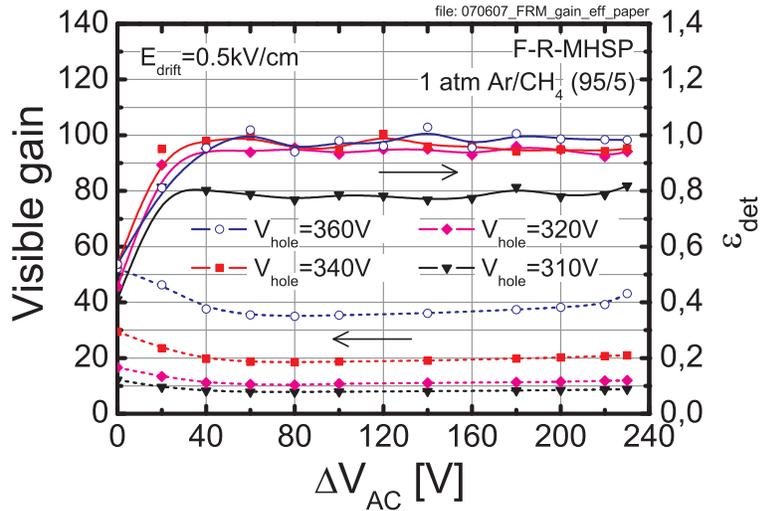}
} \caption {The visible gain (left y-axis, dashed lines) and the
photoelectron detection efficiency  (right y-axis, solid lines) as a
function of the inter-strip voltage of a F-R-MHSP in TPC conditions
(E$_{drift}$=0.2kV/cm) (a) and GPM conditions (E$_{drift}$=0.5kV/cm)
(b). Measurements performed at different values of hole voltage.}% caption for the whole figure
\label{figure:6}
\end{center}
\end{figure}

In our setup, the transfer field was set to 1kV/cm. In
\ref{figure:6}, the photoelectron detection efficiency of the
F-R-MHSP is shown for TPC conditions (E$_{drift}$=0.2 kV/cm) and for
GPM conditions (E$_{drift}$=0.5 kV/cm). In each regime, the
measurements were performed at different hole voltages of 310V,
320V, 340V and 360V.

We can see in \ref{figure:6} that:

\begin{enumerate}

\item
The visible gain of the F-R-MHSP does not depend on variations in
the inter-strip voltage. This can be considered as a first
indication of a good focusing of photoelectrons into the holes
(independent on the inter-strip voltage).
\item
The photoelectron detection efficiency is low for small inter-strip
voltages. This can be attributed to a partial collection of
photoelectrons by the narrow anode strips. As we increase the
inter-strip voltage, the efficiency is rising up, reaching a
plateau.
\item
The minimal hole-voltage which provides close to full photoelectron
detection efficiency was measured to be 320V. It corresponds to a
visible gain of $\sim$10 (\ref{figure:6}).

\end{enumerate}

In some conditions the single electron pulse height spectrum
deviates from a pure exponential, leading to an error in
normalization. This is the case for the data points in
\ref{figure:6:a}, V$_{hole}$=360V and $\Delta$V$_{AC}$<80V, where a
Polya distribution was observed, leading to overestimated efficiency
values.

\begin{figure} [h]
  \begin{center}
  \makeatletter
    \renewcommand{\p@figure}{figure\space}
  \makeatother
    \epsfig{file=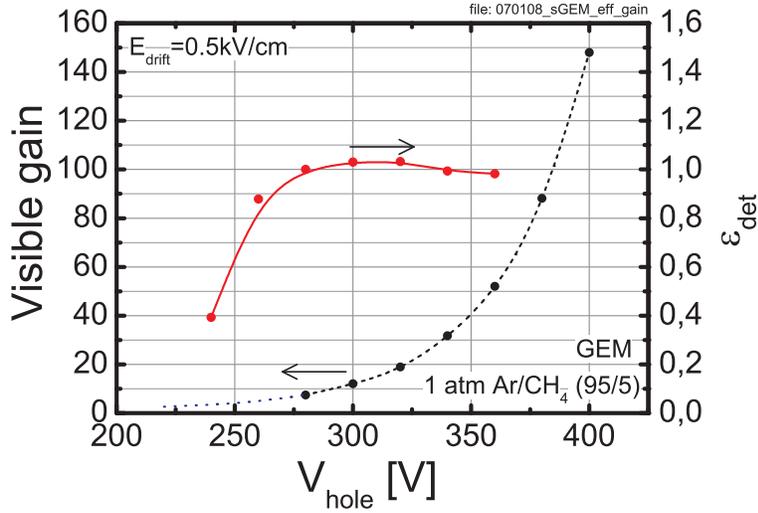, width=10cm}
    \caption{The visible gain  (left y-axis, dashed line) and the photoelectron
detection efficiency (right y-axis, solid line) as a function of the
hole voltage of the GEM in GPM conditions (E$_{drift}$ =0.5kV/cm).
For voltages below 280V (dotted line), the gain curve was
extrapolated with an exponential function (dotted line).}
    \label{figure:7}
  \end{center}
\end{figure}

\textbf{Measurements with GEM} As we mentioned above, the
photoelectron detection efficiency of the GEM was not yet measured
at a hole voltage lower than 320V \cite{richter:02}. The detection
efficiency of a single GEM as a function of its visible gain is
presented in \ref{figure:7}. During measurements, the transfer field
between the GEM and the MWPC was kept at 1kV/cm. It was found that
the minimal hole voltage which permits the operation of the GEM at
full detection efficiency for single electrons is around 280V. That
corresponds to a visible gain of $\sim$10 on the GEM.

\subsection{Studies of ion blocking capability of the first element in a cascade}

\begin{figure} [h]
  \begin{center}
  \makeatletter
    \renewcommand{\p@figure}{figure\space}
  \makeatother
    \epsfig{file=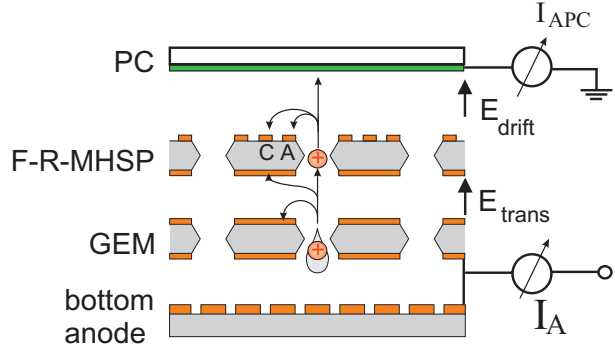, width=8cm}
    \caption{The schematic view of the setup for measurements of ion blocking
capability of the F-R-MHSP. Here the GEM serves as a source of
avalanche ions.  The avalanche charge was collected at the
interconnected GEM-bottom and bottom anode electrodes. Similar
measurements were performed with R-MHSP and GEM elements followed by
a GEM.}
    \label{figure:8}
  \end{center}
\end{figure}

\begin{figure}[!h]
\begin{center}
\renewcommand{\thesubfigure}{\thefigure\alph{subfigure}}
\makeatletter
\renewcommand{\@thesubfigure}{\Large(\alph{subfigure})}
\renewcommand{\p@subfigure}{figure\space}
\renewcommand{\p@figure}{figure\space}
\makeatother %
\subfigure [] [] % caption for subfigure a
{
    \label{figure:9:a}
    \includegraphics[width=10cm]{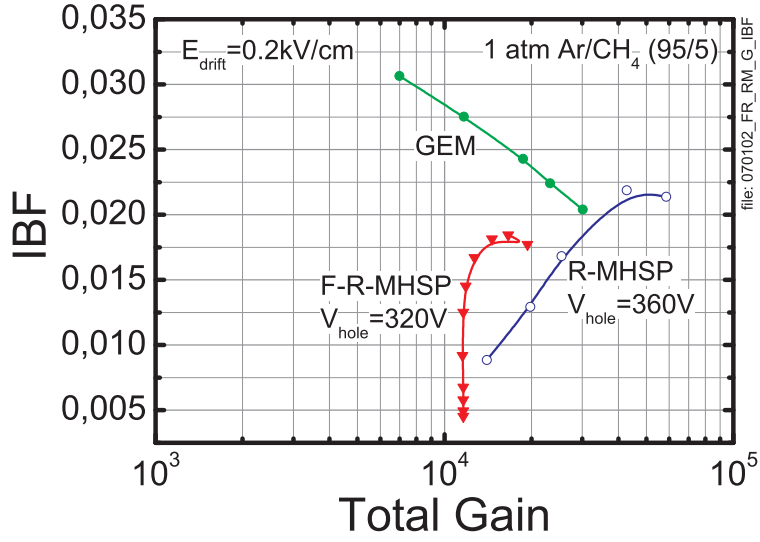}
}\hspace{1cm} %
\subfigure [] [] % caption for subfigure b
{
    \label{figure:9:b}
    \includegraphics[width=10cm]{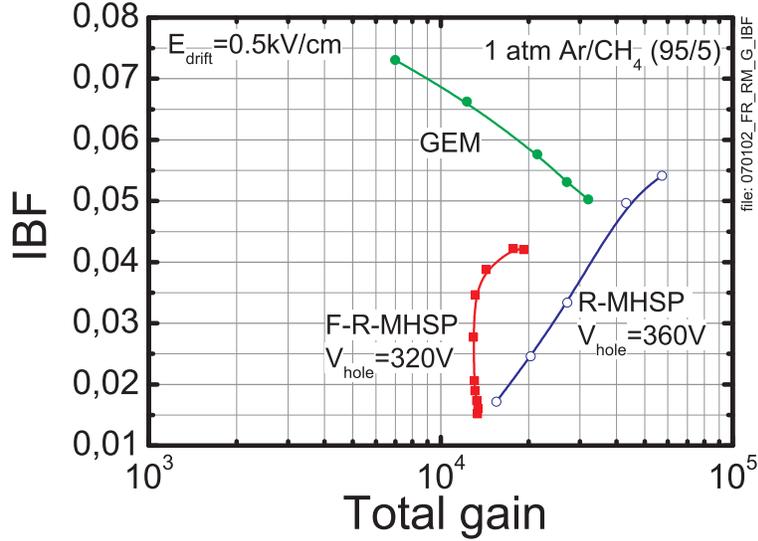}
}%
\caption {The IBF and total gain measured in the setup of \protect
\ref{figure:8} with R-MHSP/GEM, F-R-MHSP/GEM  and 2GEM
configurations, at TPC (a) and GPM (b) conditions. In the case of
R-MHSP/GEM and F-R-MHSP/GEM the gain was varied by changing the
inter-strip voltage from 0V to 60V and from 0V to 230V
correspondingly; in the case of 2GEMs - by changing the hole voltage
from 280V to 340V. The hole voltages of the R-MHSP and F-R-MHSP were
fixed at the values indicated in the figure, to ensure full photoelectron detection efficiency.}% caption for the whole figure
\label{figure:9}
\end{center}
\end{figure}

The IBF reduction capability of the first element was studied in a
setup depicted in \ref{figure:8} with a first element being a GEM, a
R-MHSP or a F-R-MHSP. It was followed by a GEM, of which the
avalanche acts as a source of back-flowing ions. This second GEM
element was biased at 420V (gain $\sim$2000); the transfer field in
the gap between the two elements was E$_{trans}$=1kV/cm and the
drift field was E$_{drift}$=0.5kV/cm (GPM conditions)
(\ref{figure:9:a}) and 0.2kV/cm (TPC conditions) (\ref{figure:9:b}).

The total avalanche current in this configuration was measured as
the sum of currents from the bottom anode and the bottom GEM
electrode (as shown in \ref{figure:8}). The IBF was calculated as
the ratio of the PC avalanche current I$_{APC}$ (under avalanche
multiplication), to the total avalanche current I$_{A}$:

\begin{equation}
\label{eq:IBF} \textrm{IBF}=\frac{\textrm{I}_{APC}}{\textrm{I}_{A}}
\end{equation}

Where the PC avalanche current I$_{APC}$ was calculated as a
difference of the total PC current under multiplication I$_{TOTPC}$
and the initial I$_{0PC}$ PC current: I$_{APC}$ = I$_{TOTPC}$ -
I$_{0PC}$.

The correlation between the IBF and the total gain (of both
elements) measured in these conditions is presented in
\ref{figure:9}. The parameters (fixed and variable) in these
measurements were the following:

\begin{itemize}
\item
R-MHSP: the inter-strips voltage ($\Delta$V$_{AC}$) varied from 0V
to 60V and the hole voltage was set to 360V (following the results
of previous section);
\item
F-R-MHSP: the inter-strips voltage ($\Delta$V$_{AC}$) varied from 0V
to 230V and the hole voltage was set to 320V (following results of
the previous section);
\item
GEM, the hole voltage (V$_{hole}$) was varied in the range
280V-340V.
\end{itemize}

In TPC conditions, the R-MHSP and the F-R-MHSP followed by a GEM
performed 3 and 6 times better, respectively, than the double-GEM in
terms of IBF reduction at a total gain of $\sim$1.2$\cdot$10$^{4}$.
The IBF values of all three multipliers improved at the lower drift
field; e.g., that of the F-R-MHSP reached the value of 0.005.

One can note that in GPM conditions, the IBF measured with either
the R-MHSP/GEM or the F-R-MHSP/GEM is 4-fold lower compared to that
of a double-GEM, at a gain of $\sim$1.5$\cdot$10$^{4}$
(\ref{figure:9}). In these conditions, both the R-MHSP and the
F-R-MHSP provided practically the same IBF values, of about 0.015.

\subsection{IBF in cascaded multipliers incorporating R-MHSP, F-R-MHSP, GEM and
MHSP elements.}

\begin{figure}[!h]
\begin{center}
\renewcommand{\thesubfigure}{\thefigure\alph{subfigure}}
\makeatletter
\renewcommand{\@thesubfigure}{\Large(\alph{subfigure})}
\renewcommand{\p@subfigure}{figure\space}
\renewcommand{\p@figure}{figure\space}
\makeatother %
\subfigure [] [] % caption for subfigure a
{
    \label{figure:10:a}
    \includegraphics[width=7cm]{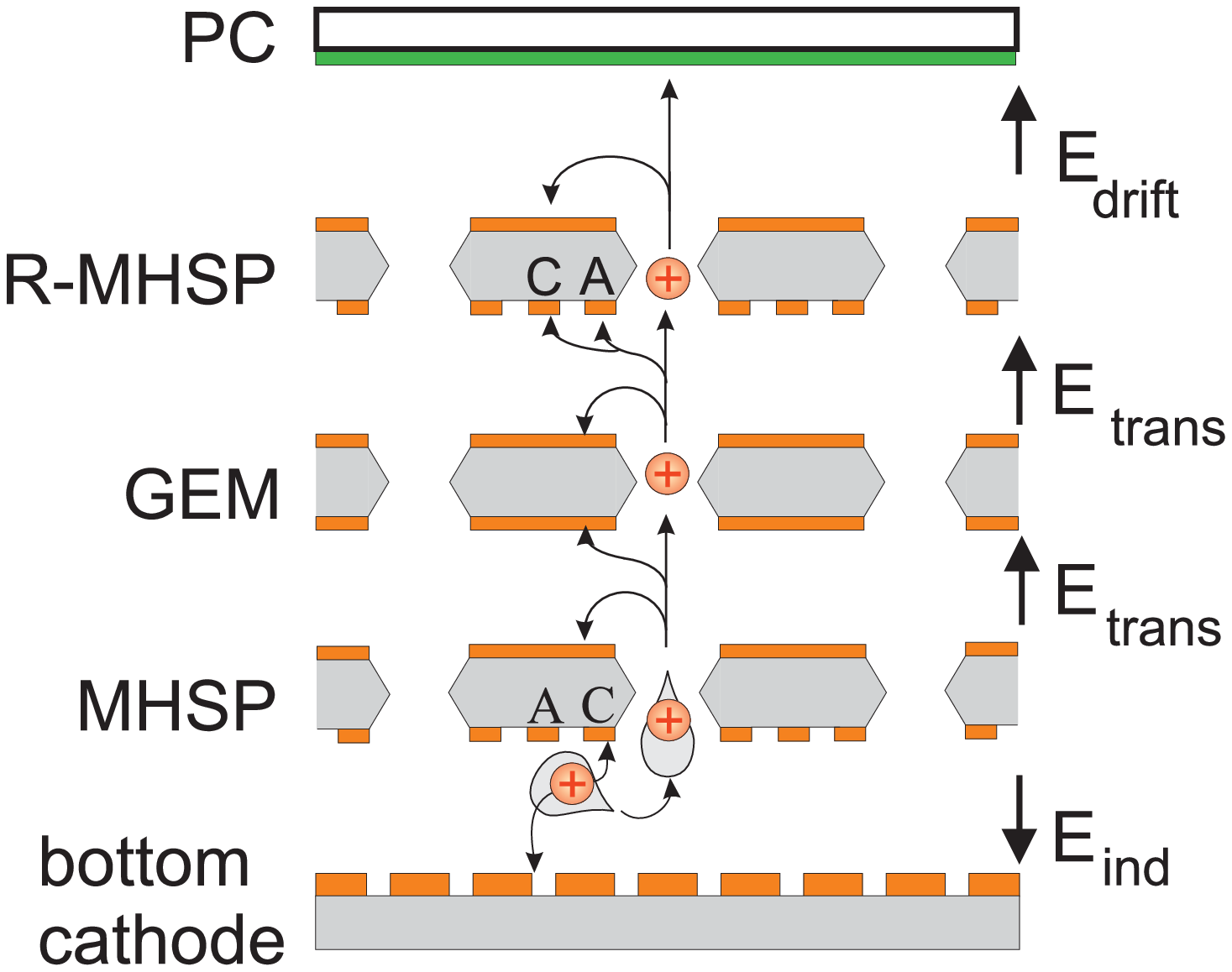}
}\hspace{0cm} %
\subfigure [] [] % caption for subfigure b
{
    \label{figure:10:b}
    \includegraphics[width=7.3cm]{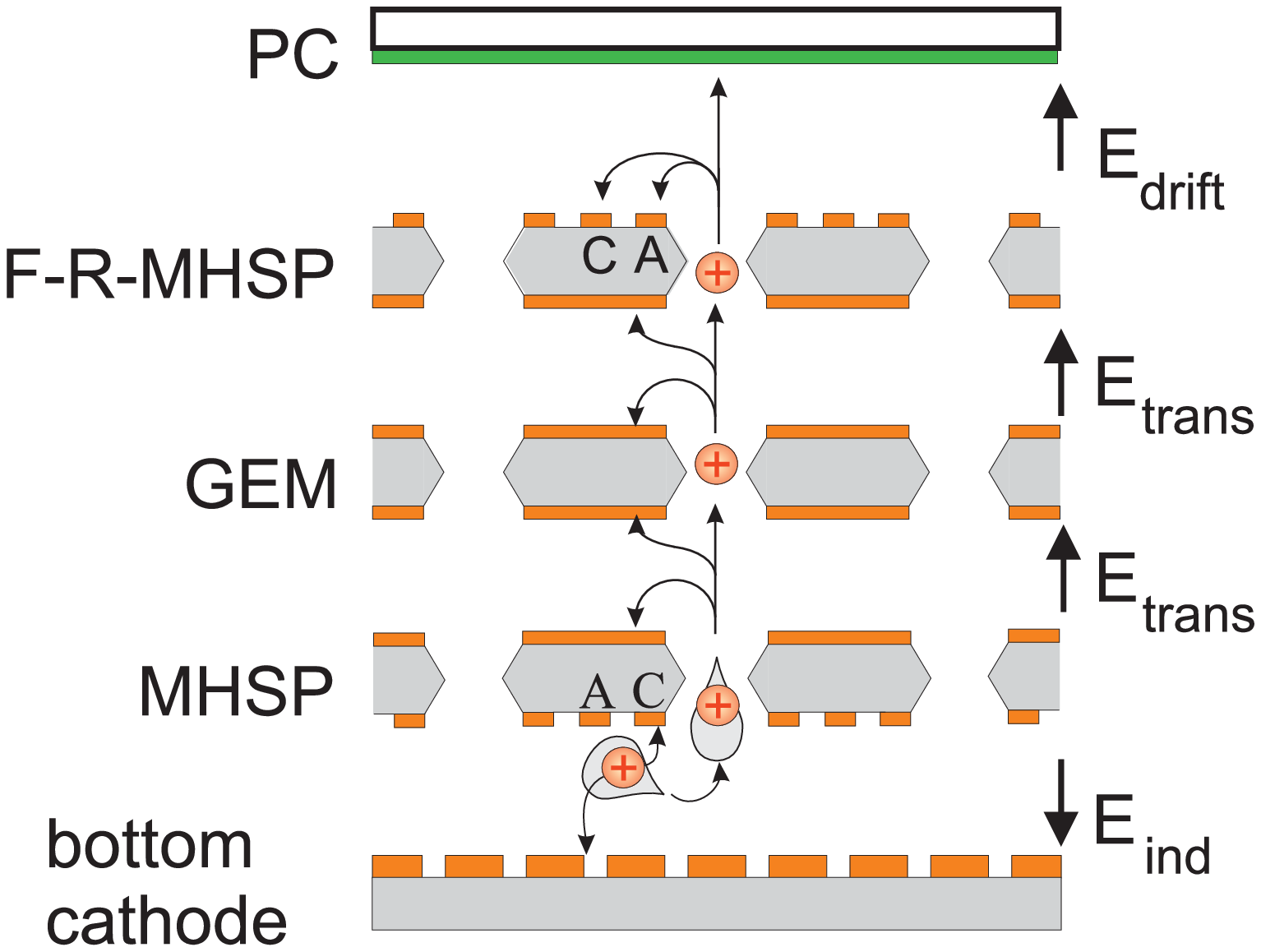}
}%
\caption { Schemes of cascaded R-MHSP/GEM/MHSP (a) and
F-R-MHSP/GEM/MHSP (b) multipliers coupled to a semi-transparent
photocathode; possible avalanche ions paths are also shown.}% caption for the whole figure
\label{figure:10}
\end{center}
\end{figure}

Systematic investigations were carried out in two types of cascades
multipliers: the R-MHSP/GEM/MHSP (\ref{figure:10:a}) and
F-R-MHSP/GEM/MHSP (\ref{figure:10:b}). The last MHSP element in each
configuration was chosen based on its known 5-fold IBF reduction
compared to a GEM \cite{maia:04}, in addition to that of the two
types of first-element multipliers investigated here. The optimized
transfer- and induction-field configurations suggested in
\cite{maia:04, killenberg:04} were combined with the insight from
the F-R-MHSP and R-MHSP studies described above. The measurements
were performed both in TPC conditions (E$_{drift}$=0.2kV/cm) and GPM
conditions (E$_{drift}$=0.5kV/cm). The following parameters were
chosen (see figure 10): E$_{trans1}$=1kV/cm; E$_{trans2}$=60V/cm
(following \cite{killenberg:04}) and E$_{ind}$=-5kV/cm; the latter
"reversed" field, permitted collecting most of the last-avalanche
ions at the bottom mesh cathode (following \cite{maia:04}). The
voltages across the holes and between strips for different elements
were the following: The first-element voltages were chosen according
to the results described above:  V$_{hole1}$=320V;
$\Delta$V$_{AC1}$=230V for the F-R-MHSP and V$_{hole1}$=360V;
$\Delta$V$_{AC1}$=60V for the R-MHSP. The GEM and MHSP potentials
were chosen as follows:
\begin{itemize}
\item
In TPC conditions, for the F-R-MHSP/GEM/MHSP detector:
V$_{GEM1}$=230V; for the R-MHSP/GEM/MHSP detector: V$_{GEM1}$=240V.
\item
In GPM conditions, we had to increase the GEM voltage and further
optimize the second transfer field. For the F-R-MHSP/GEM/MHSP:
V$_{GEM1}$=275V, E$_{trans2}$=75V/cm; for the R-MHSP/GEM/MHSP:
V$_{GEM1}$=300V, E$_{trans2}$=100V/cm.
\end{itemize}

The last-element MHSP multiplier was polarized, in both setups, as
follows: V$_{hole2}$=370V and $\Delta$V$_{AC2}$ was varied between
140V and 230V, to adjust the total gain of the whole cascaded
detector.

\begin{figure}[!h]
\begin{center}
\renewcommand{\thesubfigure}{\thefigure\alph{subfigure}}
\makeatletter
\renewcommand{\@thesubfigure}{\Large(\alph{subfigure})}
\renewcommand{\p@subfigure}{figure\space}
\renewcommand{\p@figure}{figure\space}
\makeatother %
\subfigure [] [] % caption for subfigure a
{
    \label{figure:11:a}
    \includegraphics[width=10cm]{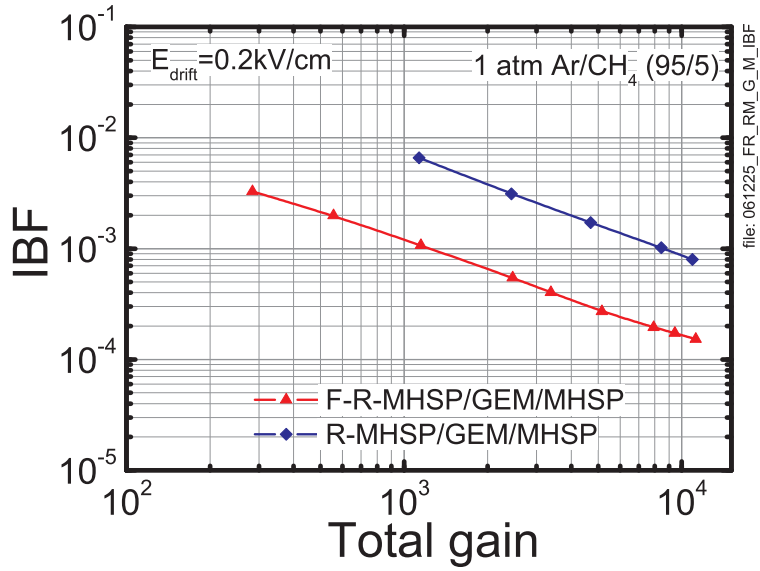}
}\hspace{1cm} %
\subfigure [] [] % caption for subfigure b
{
    \label{figure:11:b}
    \includegraphics[width=10cm]{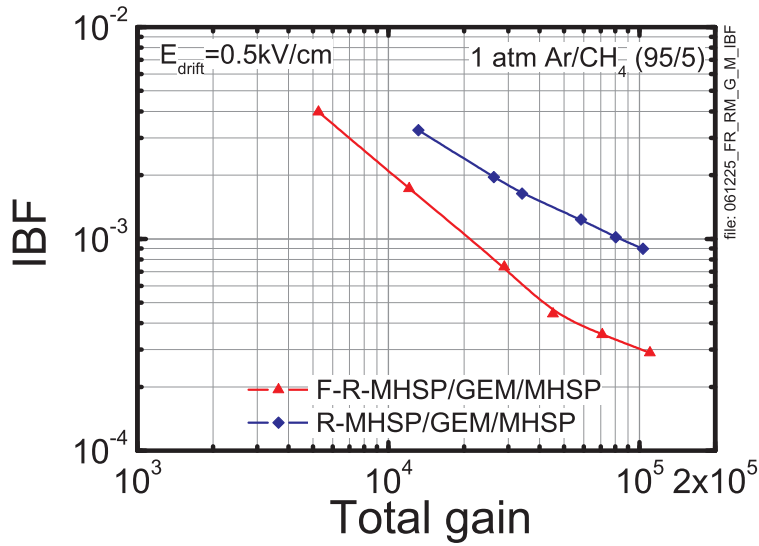}
}%
\caption {The IBF in correlation with the total gain of the
R-MHSP/GEM/MHSP (\protect \ref{figure:10:a}) and F-R-MHSP/GEM/MHSP
(\protect \ref{figure:10:b}) cascaded detectors, with
semitransparent photocathodes; the IBF is plotted for drift fields
of 0.2kV/cm (TPC conditions) (a) and 0.5kV/cm (GPM conditions) (b).}
% caption for the whole figure
\label{figure:11}
\end{center}
\end{figure}

The IBF is presented in correlation with the total gain, for
drift-field values of 0.2kV/cm (\ref{figure:11:a}) and 0.5kV/cm
(\ref{figure:11:b}).

The IBF recorded at a total gain of $\sim$10$^{4}$ in the TPC
operation mode (drift field of 0.2kV/cm) was
$\sim$1.5$\cdot$10$^{-4}$ with a F-R-MHSP first-element multiplier;
it was $\sim$8$\cdot$10$^{-4}$ with a R-MHSP first element. In these
conditions, the respective numbers of ions back-flowing into the
drift region are: $\sim$1.5 and $\sim$8 ions per primary ionization
electron.

In the GPM operation mode, the lowest IBF value reached was
3$\cdot$10$^{-4}$ for the F-R-MHSP/GEM/MHSP and 9$\cdot$10$^{-4}$
for R-MHSP/GEM/MHSP, at a detector total gain of $\sim$10$^{5}$. It
means that per single-photon event, on the average, 30 or 90 ions
reach the PC in a cascaded detector with a F-R-MHSP or a R-MHSP
first-element multiplier, correspondingly.

\section{General discussion}

In this work we have continued our long ongoing studies of IBF
reduction in cascaded electron multipliers, searching for further
improvements that will permit more stable gaseous detector
operation.

Following the 5-fold and the additional 3-fold IBF reduction
(compared to GEM) with MHSP and R-MHSP, respectively, we
investigated here a third operation mode of the MHSP element: the
"flipped Revered-bias Micro-Hole \& Strip Plate" (F-R-MHSP). Unlike
in MHSP and the R-MHSP, in the F-R-MHSP mode the strips are facing
towards the drift region. This operation mode permits uniquely
capturing both: ions originated from the first multiplier and that
induced by the avalanches of the successive cascade elements. The
F-R-MHSP is therefore best suitable as the first element of a
cascaded multiplier.

A systematic comparative study of the F-R-MHSP, R-MHSP and GEM
elements yielded operation conditions with full collection
efficiency of primary electrons into the multiplying holes and the
efficient avalanche-electrons transfer into the following elements
of a cascade. Conditions were found in which the inter-strip
potentials in the F-R-MHSP and R-MHSP were optimized for both:
electron collection and ion blocking.

It was found that for R-MHSP at hole voltages of 360V, 380V and
400V, the inter-strip potential can be raised to 60V, 70V and 90V
correspondingly, maintaining full single-electron detection
efficiency. This was measured at a fixed transfer field of 1kV/cm.
Further increase of the transfer field will allow further increase
of the inter-strip voltages; this will allow to divert more ions to
the cathode strips, with no sacrifice to the photoelectron detection
efficiency \cite{lyashenko:06}.

Field distortions in the drift region, at the hole vicinity, due to
the applied F-R-MHSP inter-strip voltage, did not affect the
electron focusing into the hole apertures under the current
operation conditions (\ref{figure:6}). This is due to the very
intense focusing field in the hole vicinity which is effectively
focusing the drifting electrons. It should be mentioned that, the
drift field is not uniform within a small region of few hundred
microns above the F-R-MHSP's top surface (\ref{figure:2:b}), in
which the back-flowing ions are trapped. The main limitation on the
detection efficiency of the F-R-MHSP arises from insufficient
multiplication within the holes; full detection efficiency, in the
present operation conditions, was reached for hole voltages
exceeding 320V at a transfer fields of 1kV/cm.

Theoretical calculations of the minimal visible gain required for
close-to-full single-photoelectron detection efficiency with R-MHSP,
performed in \cite{lyashenko:06}, were partly confirmed by the
present experimental data. It was calculated in \cite{lyashenko:06}
that a visible gain of 25 for the R-MHSP implies that, with an
exponential single-electron pulse-height distribution, in 92\% of
the events at least two electrons are transferred to the following
element in the cascade. Indeed, the present experimental
measurements showed that a single photoelectron is detected with
$\sim$100\% efficiency at a visible gain close to 20 (\ref{figure:5}
and \ref{figure:6}). Experimentally, we found single-electron
detection efficiency >90\% already at a visible gain of about 10 and
close to 100\% at a visible gain of 20 and above (\ref{figure:5} and
\ref{figure:6}).

The ion blocking capability was studied in cascaded detector
configurations where a GEM, R-MHSP or F-R-MHSP was used as the first
element, followed by a GEM; the latter served as the source of
avalanche ions.

In TPC conditions, a 3-fold lower IBF was reached with a R-MHSP
compared to that with a first-element GEM; the F-R-MHSP yielded a
6-fold better ion blocking compared to GEM.

In GPM conditions, the experiment showed 4-fold lower IBF with
either first-element R-MHSP or F-R-MHSP, compared with that of a
standard GEM. It should be noted however that it was not possible to
maintain both high gain (above 10$^{4}$) and low IBF in "GPM
conditions", for the detector comprised of only two multiplication
stages. Naturally, additional elements could be added to the cascade
to provide higher total gains.

In the F-R-MHSP/GEM/MHSP detector with a semitransparent PC, the IBF
value reached in a GPM mode (E$_{drift}$=0.5kV/cm), compatible with
full single-electron detection efficiency, was 3$\cdot$10$^{-4}$ at
a total gain of $\sim$10$^{5}$. This record IBF value seems to be
sufficient for a stable operation of the multiplier in combination
with a visible-light sensitive photocathode (e.g. bialkali). With
the resulting 30 ions impinging on the photocathode per
single-photoelectron event, a bialkali PC will induce on the average
0.6 secondary electrons (calculated according to
\cite{mormann_thes}). This could be a significant step forward in
the field of GPMs; it will permit, on one hand, a stable DC
operation of a detector with single-photon sensitivity in the
visible spectral range and on the other hand will drastically reduce
the photocathode aging \cite{breskin:05}. However, one should keep
in mind that the number of secondary electrons can fluctuate, due to
fluctuations of the number of electrons in individual avalanches;
therefore there could be more than one secondary electron per
avalanche. This phenomenon is well understood for low-pressure gases
setting the condition for breakdown. The breakdown does not occur as
far as the average number of secondary electrons liberated from the
PC by the back-flowing avalanche ions is below one
\cite{wijsman:49}. At high pressures this condition could not always
be applicable, because of possible space charge formation
\cite{loeb:40} at the PC vicinity. It is probable, however, that at
low IBF values, the space charge formation will be prevented and the
above mentioned condition for breakdown will be held.

The operation of cascaded visible-sensitive GPMs is currently under
investigation, with bialkali photocathodes coupled to
F-R-MHSP/GEM/MHSP multipliers that provided the lowest IBF values.

The IBF value of $\sim$1.5$\cdot$10$^{-4}$ measured in TPC
conditions (E$_{drift}$=0.2kV/cm) at a gain of 10$^{4}$ is also the
lowest value ever achieved. It is $\sim$30 fold lower than the
5$\cdot$10$^{-3}$ value recorded in a 3GEM TPC element at a similar
gain \cite{killenberg:04, lotze:06, roth:04}. In our lower
drift-field TPC conditions, much less ions flow back to the drift
region; indeed, the IBF decreases linearly with the drift field
\cite{bondar:03}. With a first-element F-R-MHSP less than 2 ions per
avalanche electron return on the average to the drift region; with a
first-element R-MHSP, 8 ions flow back to the drift region.

These very low numbers could be of prime importance for the
conception of TPC detectors operating in DC mode, with large
particle multiplicities at high repetition rates.

Further reduction of the IBF could be reached naturally with
additional patterned hole multipliers in the cascade. We are
presently investigating the idea of double-face patterned
hole-multipliers, with ion-defocusing strips running on both faces

\acknowledgments

This work is partly supported by the Israel Science Foundation,
grant No 402/05, by the MINERVA Foundation and by Project
POCTI/FP/63441/2005 through FEDER and FCT (Lisbon). A. Breskin is
the W.P. Reuther Professor of Research in The Peaceful Use of Atomic
Energy.

\end{document}